\documentclass[article]{aa}  
\usepackage{graphicx}
\usepackage{txfonts}
\usepackage{natbib}
\bibpunct{(}{)}{;}{a}{}{,}
\begin{document}
   \title{New Baade-Wesselink distances and radii for four metal-rich Galactic 
   Cepheids\thanks{Based on observations made with MPG/ESO 2.2m telescope at 
   La Silla Observatory under proposal IDs: 75.D-0676, 60.A-9120 and 
   multi-epoch, multi-band NIR data at SAAO.}}


   \author{S.\ Pedicelli\inst{1,2}
	  \and B.\ Lemasle\inst{3}
	  \and M.\ Groenewegen \inst{4}
	  \and M.\ Romaniello \inst{1}
	  \and G.\ Bono\inst{2,5}
	  \and C. D.\ Laney,\inst{6}
	  \and P.\ Fran\c{c}ois \inst{7}
	  \and R.\ Buonanno,\inst{2,8}
	  \and F.\ Caputo,\inst{5}
	  \and J.\ Lub,\inst{9}
	  \and J.\ W.\ Pel,\inst{3}
	  \and F.\ Primas \inst{1}
	  \and J.\ Pritchard\inst{1}
}
   \institute{European Southern Observatory (ESO) Karl-Schwarzschild-Strasse 2, D-85748 Garching bei M{\"{u}}nchen, Germany; \email{spedicel@eso.org}
    \and Universit\`a di Roma Tor Vergata, Via della Ricerca Scientifica 1, 00133 Roma, Italy
    \and Kapteyn Institute, University of Groningen, P.O. Box 800, 9700 AV Groningen, The Netherlands
    \and Royal Observatory of Belgium, Ringlaan 3  B-1180 Brussels, Belgium
    \and INAF -- Osservatorio Astronomico di Roma, Via Frascati 33, Monte Porzio Catone, Italy
    \and South African Astronomical Observatory, PO Box 9, 7935 Observatory, South Africa
    \and Observatoire de Paris-Meudon, GEPI, 61 avenue de l'Observatoire, F-75014 Paris, France
    \and ASI--Science Data Center, ASDC c/o ESRIN, via G. Galilei, 00044 Frascati, Italy 
    \and Leiden Observatory, Leiden University, P.O. Box 9513, NL-2300 RA  Leiden, The Netherlands
}

  \date{Received September 15, 1996; accepted March 16, 1997}

 
  \abstract
    {}
   {We provided accurate estimates of distances, radii and iron abundances  
   for four metal-rich Cepheids, namely V340~Ara, UZ~Sct, AV~Sgr and VY~Sgr.
   The main aim of this investigation is to constrain their pulsation properties 
   and their location across the Galactic inner disk.}
   {We adopted new accurate NIR (J,H,K) light curves and new radial velocity measurements 
    for the target Cepheids to determinate their distances and radii using the Baade-Wesselink 
    technique. In particular, we adopted the most recent calibration of the IR surface brightness 
    relation and of the projection factor. Moreover, we also provided accurate measurements of the 
    iron abundance of the target Cepheids.}
   {Current distance estimates agree within one $\sigma$ with similar distances based either on empirical 
    or on theoretical NIR Period-Luminosity relations. However, the uncertainties of the Baade-Wesselink 
    distances are on average a factor of 3-4 smaller when compared with errors affecting other distance 
    determinations. Mean Baade-Wesselink radii also agree at one $\sigma$ level with Cepheid radii 
    based either on  empirical or on theoretical Period-Radius relations. Iron abundances are, 
    within one $\sigma$, similar to the iron contents 
    provided by Andrievsky and collaborators, thus confirming the super metal-rich nature of the target 
    Cepheids. We also found that the luminosity amplitudes of classical Cepheids, at odds with RR Lyrae 
    stars, do not show a clear correlation with the metal-content. This circumstantial evidence appears 
    to be the consequence of the Hertzsprung progression together with the dependence of the topology 
    of the instability strip on metallicity, evolutionary effects and binaries. 
    }
   {}


   \maketitle
%

\section{Introduction}
Classical Cepheids are used both as standard candles and tracers 
of young stellar populations \citep{mac09, mio09}. 
They are bright and variable objects and thanks 
to the Hubble Space Telescope they have been identified 
and accurately measured in Local Group (d $\lesssim$ 1 Mpc) and 
in Local Volume (d $\lesssim$ 10 Mpc) galaxies \citep{free01, tamm03, bo08}. 
They obey to a Period-Luminosity (PL) relation and are the most popular primary 
distance indicators \citep{feast99, macr06, fo07, groe07, diben08, groe08,ker08,
kan09a, mar09, sco09}. In spite of these outstanding 
observational and theoretical efforts \citep[][and references therein]{marc09} 
the universality of optical and near-infrared (NIR) PL relations still 
lacks a firm empirical validation \citep{ben07, ro08, sand09}.
   
This thorny problem remains even if the zero-point and the slope 
of the PL relation can be estimated with a variety of independent 
methods. Ideally the calibration and the validation of the PL relation 
should be rooted on distances measured with a geometrical method such 
as the trigonometric parallaxes. This approach was recently adopted by 
\cite{ben07} who provided parallaxes with a mean accuracy of 8\% for a 
sample of nine Galactic Cepheids using the {\em Fine Guidance Sensor} (FGS) on 
board of the Hubble Space Telescope (HST). A new revision of HIPPARCOS parallaxes for Galactic Cepheids
(244 objects) has been recently provided by \cite{vanl07} and by \cite{vanletal07}.  
The accuracy of the new measurements is on average a factor of two 
better than the old ones. However, the Cepheids with the most accurate 
HIPPARCOS parallaxes are Polaris ($\alpha$ UMa) and the prototype 
$\delta$ Cep. The accuracy of the former one is 1.6\% \citep{vanletal07}, 
while the latter is similar to the accuracy of the FGS@HST parallax, 
namely  5.2\% versus 4.1\% \citep{mer05}.     

The Baade-Wesselink method (BW, \cite{ba26, wess46}) provides an independent 
empirical approach to measure Cepheid absolute distances and it can be applied 
to variable stars. This method relies on two observables: the radial velocity 
($v_r$) and the variation of the angular diameter ($\theta$). The latter parameter 
was historically substituted by the variation in color along the pulsation cycle.  
However, direct measurements of the Cepheid angular diameter have been 
recently provided by \cite{ker04b} using the Very Large Telescope Interferometer
(VLTI). In particular, the use of VLT INterferometer Commissioning Instrument (VINCI) 
gave the unique opportunity to provide angular diameter measurements along the 
pulsation cycle for seven Cepheids. These pioneer measurements encouraged 
a detailed comparison between theory and observations concerning the 
limb darkening and the atmosphere of variable stars \citep{mar03,nar07}, 
but the number of Cepheids for which these measurements are available is still 
very limited. 

\begin{table*}\label{mag}
\caption{Intrinsic parameters, mean NIR magnitudes and mean radial velocities for the 
target Cepheids.} 
\begin{center}
\begin{tabular}{l c c c c c c c c c c} \hline\hline
NAME    &$\alpha$(J2000)$^a$&$\delta$(J2000)$^a$&$\log$ P&E(B-V)$^b$&$<J>^c\pm \sigma(J)$&$<H>^c\pm \sigma(H)$&$<K>^c\pm\sigma(K)$&$N_s^d$&$<v_r>^e$&$\Delta v_r^e$\\ \hline
V340~Ara& 16 45 19&-51 20 33&1.32&0.574&7.382$\pm$0.011&6.809$\pm$0.007&6.619$\pm$0.008&25+2&-80.8$\pm$1.2&59.1 \\
UZ~Sct  & 18 31 22&-12 55 00&1.17&1.071&7.502$\pm$0.049&6.818$\pm$0.042&6.564$\pm$0.045&25+2& 40.2$\pm$0.5&49.4 \\
AV~Sgr  & 18 04 49&-22 43 00&1.19&1.267&6.909$\pm$0.040&6.081$\pm$0.035&5.758$\pm$0.033&25+1& 19.3$\pm$1.8&59.3 \\
VY~Sgr  & 18 12 05&-20 42 00&1.13&1.283&7.174$\pm$0.069&6.375$\pm$0.046&6.068$\pm$0.040&25+2& 16.0$\pm$1.8&59.2 \\
\hline
\multicolumn{11}{l}{$^a$ Cepheid coordinates: units of right ascension are   
hours, minutes, and seconds; units of declination are degrees, arcminutes and arcseconds.} \\
\multicolumn{11}{l}{$^b$ Reddening according to \cite{fer95}.} \\
\multicolumn{11}{l}{$^c$ Mean NIR magnitudes.} \\
\multicolumn{11}{l}{$^d$ Number of spectra collected with FEROS at the 2.2m MPG/ESO telescope.} \\
\multicolumn{11}{l}{$^e$ Mean radial velocity and velocity amplitude (km s$^{-1}$).} \\
\end{tabular}
\end{center} 
\end{table*} 
To overcome the difficulties in the interferometric measurement of the 
angular diameter several variants of the BW method were suggested 
in the literature. Among them the methods based on the Surface-Brightness (SB) 
relations link variations in color with variations in angular diameters. 
This method can be applied to Cepheids for which accurate radial velocities 
and multi-wavelength light curves are available. \cite{sto04}, 
\cite{ker04a} and \cite{groe07} (hereafter G07) derived such relations, 
using an optical-NIR ($V$-$K$) color (IRSB), which gives the highest 
precision in the derived quantities.

However, the most relevant limit of the currently adopted BW methods is the value 
of the projection factor ($p$-factor). This parameter links radial velocity changes 
to radius changes and still lacks firm theoretical and empirical 
constraints \citep{nar09}. It can be empirically estimated using Cepheids for 
which  accurate interferometric angular diameter measurements, radial velocity 
curves and trigonometric parallaxes are available. This approach was 
applied to $\delta$ Cep by \cite{mer05} using the new optical interferometric 
measurements obtained with the CHARA Array, the FGS@HST trigonometric 
parallax by \cite{ben07} and the radial velocities available 
in the literature. The $p$-factor they found --$p$=$1.27\pm0.06$-- agrees 
quite well with theoretical predictions by \cite{nar04b}. More recently, 
G07 found that the use of a constant $p$-factor ($p$=$1.27\pm0.05$) for six Galactic 
Cepheids, with interferometrically measured angular diameter variations and 
known distances, agrees quite well with HST parallaxes. Moreover, he found that a 
strong period dependence of the $p$-factor ($p \sim -0.15\cdot \log P$, \cite{gie05}), 
could also be ruled out. However, a moderate period dependence ($p \sim -0.03 \cdot \log P$) 
as suggested either by  \cite{gie93, gie97,gie98, bar03, sto04}, or more recently by 
\cite{nar09} ($p \sim -0.08 \cdot \log P$) is still consistent with currently available 
data (G07).

In this investigation we plan to apply the BW method using the most recent 
calibration of the IRSB relation (Groenewegen 2010, hereinafter G10, in preparation), 
to estimate 
the distance of four metal-rich Galactic Cepheids. In particular, we plan to use 
new radial velocity measurements, new accurate NIR ($J$,$H$,$K$) light curves and $V$-band 
light curves available in the literature \citep{ber92}. Moreover, we discuss in 
\S 4 the iron abundance of the four target Cepheids using \ion{Fe}{i} and \ion{Fe}{ii} lines. 
\S 5 deals with the pulsation amplitude of metal-rich Cepheids, while 
in \S 6 we summarize current findings and outline future developments 
of this project.  

\begin{figure}
\begin{center}
\includegraphics[height=10.5truecm, width=7.25cm,angle=0]{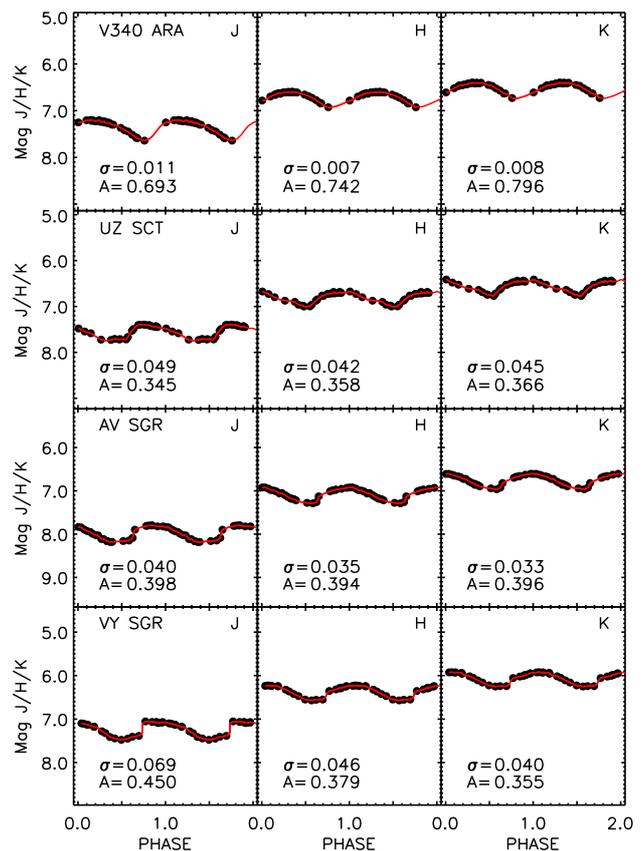} 
\vspace{0.8cm} 
\caption{From left to right $J$,$H$,$K$-band light curves for the four metal-rich 
         selected Cepheids. In each panel are also plotted the intrinsic scatter 
        ($\sigma$) of the fit with a cubic spline (red line) and the luminosity  
        amplitude.}\label{LCJHK}
\end{center}
\end{figure}

\section{Photometric and spectroscopic data}

The selected stars are the most metal-rich Cepheids in the large sample 
(113 objects) collected by \cite{and02b, and02a, and02c, and04}. Positional and 
physical parameters for the selected Cepheids are listed in Table \ref{mag}.  

\subsection{Optical and NIR data}
In order to provide accurate estimates of their distances we plan to use 
the most recent calibration of the IRSB relation \citep[][G10]{barev76, last95, fogi97, sto04}. 
To accomplish this goal one of 
us (CDL) collected accurate multi-epoch, multi-band NIR data at SAAO for 
the four targets. The typical uncertainty of individual phase points ranges 
from $0.005$ to $0.007$ mag for $K<6$ mag, deteriorating to about $0.012$ at 
$K=8.6$ mag. Figure \ref{LCJHK} shows the $J$, $H$ and $K$-band light curves 
for the target Cepheids. The intrinsic scatter ($\sigma$) of the fit with a 
cubic spline (red line) and the luminosity amplitude are also labeled. 
The mean NIR ($J$,$H$,$K$) magnitudes have been estimated as a time average 
on magnitudes and are listed in Table \ref{mag}. The 
typical uncertainties are of the order of a few hundredths of a magnitude. 

For three out of the four targets (AV~Sgr, UZ~Sct, VY~Sgr) we have 
detailed optical light curves collected in the Walraven bands 
\citep{wal64,lubpel75, pel76,lubpel77}. 
However, for the Baade-Wesselink solution we decided to use more recent $V$-band 
light curves available in the literature \citep{ber92}, since the accuracy improves 
if photometric and radial velocity data are collected close in time. 
The large sample of Cepheids with Walraven 
photometry will be used in Section 5, where we discuss the period-amplitude diagram.
%
%

\begin{figure}
\begin{center}
\includegraphics[width=6.0cm,angle=0]{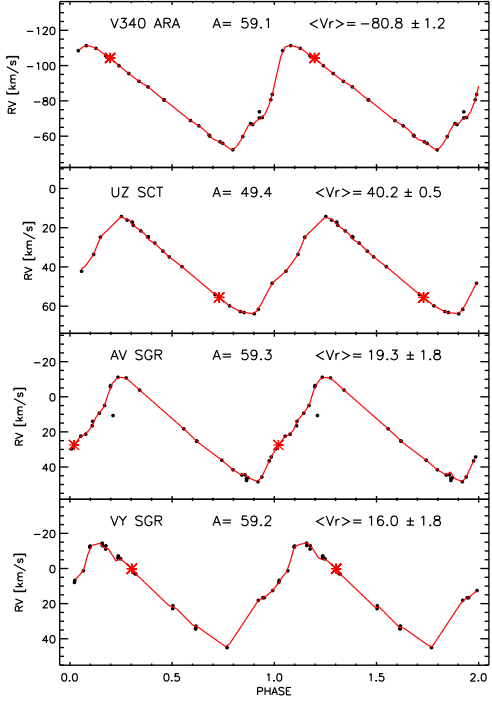}
\vspace{0.5cm} 
\caption{Radial velocity curves for the four selected Cepheids, measured by    
         cross-correlating a line list (see text). 
	 The red asterisks display the radial velocities based on 
         the high S/N spectra adopted for measuring the iron abundance. 
         The amplitude and the mean radial velocity (kms$^{-1}$) are labeled.}\label{rv}
\end{center}
\end{figure}

\begin{table}
\caption{New weak \ion{Fe}{ii} line list added to the \ion{Fe}{i} and 
\ion{Fe}{ii} line list of \cite{ro08}.} 
\begin{center}
\begin{tabular}{cccc} \hline\hline
$\lambda[\AA]$&$Ion$&$EP$& $\log~g_f$ \\ \hline
 4893.82 &  \ion{Fe}{ii} &  2.83  &    -4.45\\
 4923.93 &  \ion{Fe}{ii} &  2.88  &    -1.35\\
 4993.35 &  \ion{Fe}{ii} &  2.81  &    -3.56\\
 5100.66 &  \ion{Fe}{ii} &  2.81  &    -4.16\\
 5132.67 &  \ion{Fe}{ii} &  2.81  &    -3.95\\
 5197.58 &  \ion{Fe}{ii} &  3.23  &    -2.23\\
 5234.63 &  \ion{Fe}{ii} &  3.22  &    -2.22\\
 5256.94 &  \ion{Fe}{ii} &  2.89  &    -4.25\\
 5325.56 &  \ion{Fe}{ii} &  3.22  &    -3.18\\
 5414.07 &  \ion{Fe}{ii} &  3.22  &    -3.54\\
 5425.26 &  \ion{Fe}{ii} &   3.2  &    -3.27\\
 5534.85 &  \ion{Fe}{ii} &  3.24  &    -2.75\\
 5932.06 &  \ion{Fe}{ii} &   3.2  &    -4.81\\
 7135.02 &  \ion{Fe}{ii} &  6.21  &    -2.60\\
 7301.56 &  \ion{Fe}{ii} &  3.89  &    -3.68\\
 7310.22 &  \ion{Fe}{ii} &  3.89  &    -3.36\\
 7672.37 &  \ion{Fe}{ii} &  3.15  &    -5.19\\
 7777.11 &  \ion{Fe}{ii} &  3.20  &    -5.24\\
 7801.24 &  \ion{Fe}{ii} &  5.91  &    -2.94\\
 7841.39 &  \ion{Fe}{ii} &  3.90  &    -3.72\\
 8250.34 &  \ion{Fe}{ii} &  5.22  &    -3.29\\
 8264.72 &  \ion{Fe}{ii} &  6.81  &    -2.08\\
 8324.96 &  \ion{Fe}{ii} &  6.22  &    -2.75\\
 8330.59 &  \ion{Fe}{ii} &  6.22  &    -2.35\\
 8490.08 &  \ion{Fe}{ii} &  9.74  &     0.34\\
 8981.13 &  \ion{Fe}{ii} &  6.72  &    -2.18\\
 8990.39 &  \ion{Fe}{ii} &  6.23  &    -2.38\\
 9095.11 &  \ion{Fe}{ii} &  9.65  &     0.27\\
 9122.92 &  \ion{Fe}{ii} &  9.85  &     0.53\\
 9132.37 &  \ion{Fe}{ii} &  9.85  &     0.51\\
 9187.16 &  \ion{Fe}{ii} &  9.70  &     0.25\\
 9244.74 &  \ion{Fe}{ii} &  6.22  &    -2.40\\
 9297.27 &  \ion{Fe}{ii} &  9.65  &     0.41\\
 9464.88 &  \ion{Fe}{ii} &  6.09  &    -2.54\\
 9572.60 &  \ion{Fe}{ii} &  5.82  &    -2.87\\
 9843.19 &  \ion{Fe}{ii} &  6.14  &    -2.60\\
 9849.74 &  \ion{Fe}{ii} &  6.73  &    -2.30\\
 9956.31 &  \ion{Fe}{ii} &  5.48  &    -2.95\\\hline
\label{lines}
\end{tabular}
\end{center} 
\end{table} 

\subsection{Spectroscopic data}
Together with photometric data we also secured for the target Cepheids accurate 
multi-epoch, high-resolution (R$\sim$30,000) spectra covering the entire pulsational    
cycle with FEROS@2.2m MPG/ESO telescope.  The goal was to provide individual radial 
velocity measurements with an accuracy of 1 km/s for $\approx$ two dozen of epochs.  
The number of epochs is a strong requirement to reach with BW methods the nominal 
accuracy of 5\% in distance. The spectroscopic sample, indeed, includes more than 
25 spectra per Cepheid (see column 9 in Table \ref{mag}). These spectra were collected 
between May and September 2005. The typical exposure time is $\sim 320$s and the quality 
of the individual spectra is quite good (signal-to-noise ratio, $S/N \sim 70$) and allowed 
us to reach the required accuracy in radial velocity measurements. 
Radial velocity measurements were performed using {\tt fitline} a semi-interactive routine 
developed by one of us (PF, \cite{fran07}).  The algorithm adopted in {\tt fitline} is based on a 
cross--correlation between the lines of each spectrum and a given line list. 
The reader interested in a more detailed description of {\tt fitline} is referred to 
\cite{bert07, bert08}. Note that to properly deal with metal-rich Cepheids the iron line 
list provided by \cite{ro08} was supplemented with three dozen of weak $\ion{Fe}{ii}$ lines 
(see Table \ref{lines}).  
Figure \ref{rv} shows the individual radial velocity measurements and the fit with a 
cubic spline (red lines). The mean radial velocities ($<v_r>$) and the velocity 
amplitudes ($\Delta v_r$) based on the spline fit are also labeled (see also columns 
10 and 11 of Table \ref{mag}). 

\begin{figure}
\begin{center}
\includegraphics[height=10.0truecm, width=7.75cm,angle=0]{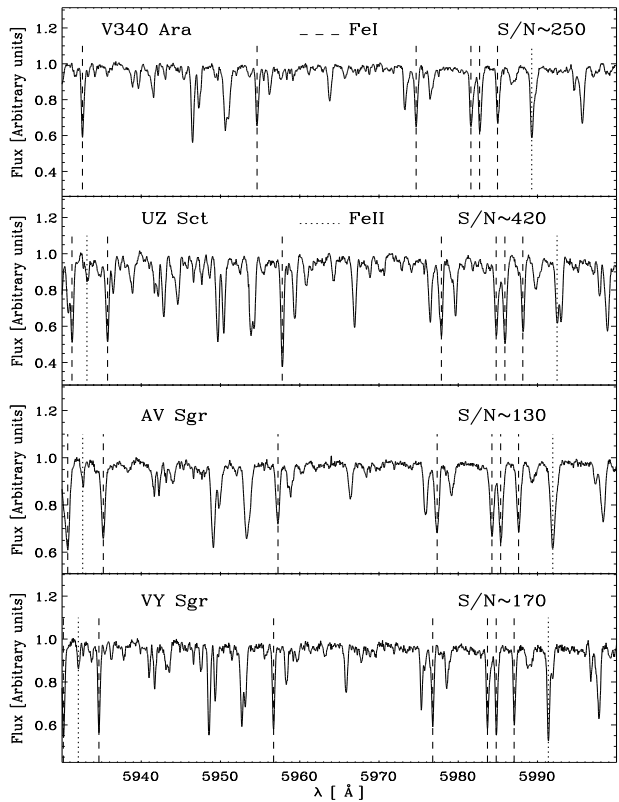}
\vspace{0.5cm} 
\caption{High S/N spectra for the four selected Cepheids in a limited wavelength 
range (5930-6000 $\AA$). Vertical dashed and dotted lines mark $\ion{Fe}{i}$ 
and $\ion{Fe}{ii}$ lines. In each panel is also plotted the S/N ratio of either 
individual (AV Sgr) or co-added spectra.}\label{spec}
   \end{center}
   \end{figure}

The target Cepheids are among the most metal-rich Galactic Cepheids and, as expected, 
they are located close to the inner edge of the thin disk \citep{mio09}. In order to 
provide accurate iron abundance measurements, we also collected for each target 1-2 high 
S/N ratio spectra with FEROS@2.2m MPG/ESO telescope (see red asterisks in Figure \ref{rv} 
and Table \ref{mag}). These spectra were collected by one of us (PF) at the end of 
March 2007. The typical exposure time is $\sim$ 1800s  and their $S/N$ ratio ranges from 
150 to 300. The abundance analysis of the Cepheids with two high $S/N$ spectra was 
performed on the co-added spectrum, since they were collected close in time. 
Figure \ref{spec} shows the co-added spectra for the four selected Cepheids in the 
wavelength range between 5930 and 6000 $\AA$. The vertical dashed and dotted lines 
mark $\ion{Fe}{i}$ ($\lambda$=5930.17, 5934.66, 5956.70, 5976.78, 5983.69, 5984.79, 
5987.05 $\AA$) and $\ion{Fe}{ii}$  ($\lambda$=5932.06, 5991.37 $\AA$) lines, respectively. 
The S/N ratio of either individual (AV~Sgr) or co-added spectra is also labeled.

\section{BW method} 
The BW method adopted to estimate distances and radii has already 
been discussed in previous papers by Groenewegen \citep{groe04, groe07, groe08}. 
In the following we briefly mention the key points of the method and the 
differences with the original approach. 
According to the definition of quasi-monochromatic surface brightness, 
we can write $M_V$ - $S_V$ + 5$\times \log (R/R_{\odot})$ = const.
The absolute magnitude, and in turn the distance, can be found by differentiating 
this equation with respect to the pulsation phase, by multiplying the result for 
a color index --$(V-K)_0$-- and then by integrating over the entire pulsation 
cycle. The radial velocity is tightly connected with the pulsation velocity
and the curve can be integrated to obtain the radius variation as a function 
of time (phase): 

{\small
\begin{equation}\label{bw}
\Delta R (t,\delta\theta) =-p \int^{t+P\delta\theta}_{t_0}{(v_r(t) - v_{\gamma}) dt}
\end{equation}
}
where $P$ is the pulsation period, $p$ the projection factor, 
$v_{\gamma}$ the systematic radial velocity and $\delta\theta$ accounts for 
a phase shift between the radial velocity curve and the angular diameter changes 
measured either interferometrically or via the SB relation.  
Eventually, we end up with the equation 
$\Theta(t)$= const.$\times$ [$R_0$ + $\Delta R(t,\delta\theta$)]/$d$  
where $\Theta(t)$ is the angular diameter in mas, $R_0$ is stellar radius in 
solar radii and $d$ the distance in parsec. 
We measure the apparent radial velocity $v_r$, i.e. the Doppler shift 
of absorption lines in the stellar atmosphere, projected along the line of 
sight and integrated over the stellar disk. To obtain the pulsational velocity 
we use the $p$-factor (see column 2 in Table \ref{distances}) 
coming from \cite{nar09}:
{\small
\begin{equation}\label{pfactor}
p=1.31\pm0.06-0.08\pm0.05 \cdot \log(P)
\end{equation} 	
}
Integration of the pulsational velocity over the entire pulsation cycle provides an 
estimate of the linear radius variation.
The angular diameter variation was derived using the IRSB method and the 
$(V - K)$ color. The calibration adopted in this investigation was derived 
by G10:
{\small
\begin{equation}\label{irsb_mart}
\log \theta_0 = 0.2692\;(V -K)_0 + 0.5298
\end{equation} 	
}
The above relation together with the Eq. \ref{bw} gives, once integrated over the 
pulsation cycle, individual Cepheid distances.   



\begin{figure*}
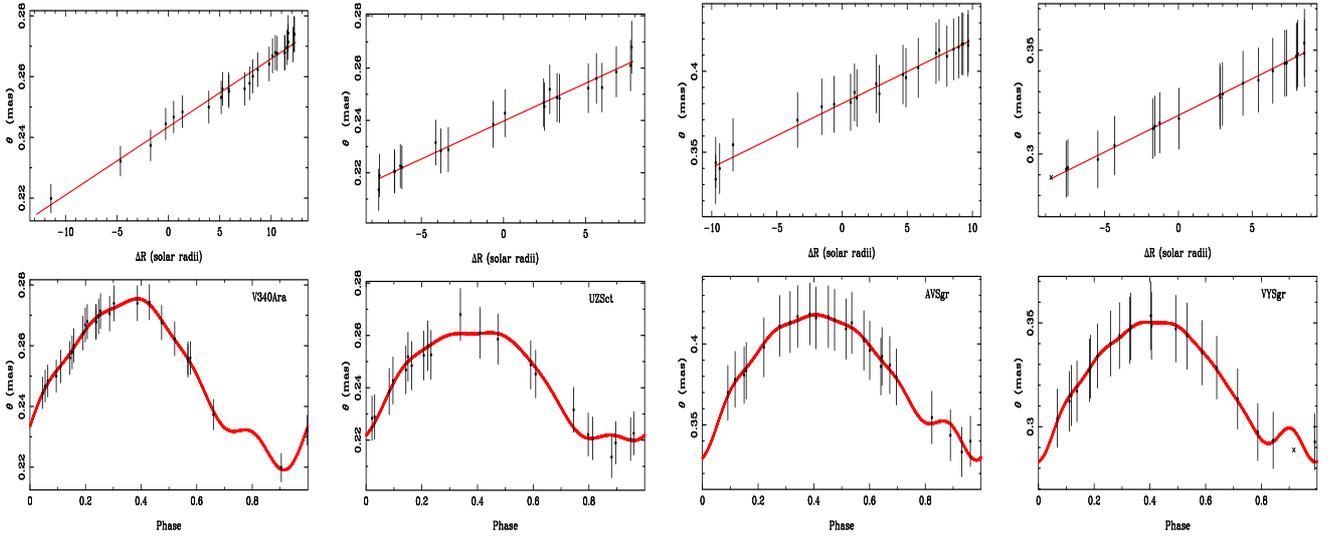

\begin{center}
\begin{tabular}{cccc}
      \resizebox{40mm}{70mm}{\includegraphics{V340Ara_BW.ps}} &
      \resizebox{40mm}{70mm}{\includegraphics{UZSct_BW.ps}} &
      \resizebox{40mm}{70mm}{\includegraphics{AVSgr_BW.ps}} &
      \resizebox{40mm}{70mm}{\includegraphics{VYSgr_BW.ps}} \\
   \end{tabular}
\caption{From left to right: V340~Ara, UZ~Sct, AV~Sgr and VY~Sgr. 
For each star, the top panels show the linear-bisector 
fit to the angular diameter as a function of radial displacement. The bottom panels 
show the angular diameter against phase. Crosses mark data-points not included in 
the fit.}\label{BW_fig}
   \end{center}
   \end{figure*}

\subsection{Cepheid distances}
Figure \ref{BW_fig} shows from left to right the application of the 
BW method to V340~Ara, UZ~Sct, AV~Sgr and VY~Sgr. For each Cepheid 
we plot the variation of the angular diameter against phase (bottom panels) 
and the change in angular diameter derived from the IRSB relation 
(see Eq. \ref{irsb_mart}) against the change in radius (top panels) 
obtained by integration of the $RV$ curve (see Eq. \ref{bw}). 
The absolute magnitudes have been derived in a self-consistent way from 
the Fourier fit to the data. Note that the fit in the phase interval 0.8-1.0 
is often poor. This limitation of the BW method was known before \citep{sto04}. 
It was argued that non-LTE effects and an increase in the micro turbulence 
during these phases may change the atmospheric structure, and in turn 
hamper the use of a simple surface-brightness relation \citep{ber97}.
Table \ref{distances} lists the estimated 
distances and radii and the comparison with previous estimates. 
The errors listed in columns 3 and 4 of Table \ref{distances} give the 
uncertainties (one $\sigma$) on distance and radius for the target Cepheids. 
They account not only for the errors in the fit, but also for the errors 
estimated running several Monte Carlo simulations by artificially changing 
individual errors on $V$, $K$ and $RV$ measurements. In order to compare 
the above distances with similar estimates available in the literature we 
adopted the $J$, $H$ and $K$-band PL relations for Galactic Cepheids recently 
provided by \cite{fo07}. The mean of the three individual distances 
together with their errors for the target Cepheids are listed in column 5 of 
Table \ref{distances}. The errors account for the uncertainty on the mean 
magnitudes and on the PL relations. The comparison indicates that distances 
based on the BW method and on NIR PL relation agree within 1$\sigma$. 
Current uncertainties on the BW distances, by taking into account the 
errors in the fit and the errors estimated using the Monte Carlo simulations, 
range from 6\% (UZ Sct, AV Sgr, VY Sgr) to 14\% (V340 Ara), while the 
uncertainties on the distances based on the empirical NIR PL relations 
range from 22\% to 34\%. 

The NIR PL relations provided by \cite{fo07} do not account for a 
possible dependence of the NIR PL relation on the metal content. We are 
dealing with Cepheids characterized by super-solar iron abundances. 
Therefore, we estimated the Cepheid distances using theoretical NIR 
PL relations including both metal-intermediate and metal-rich 
Cepheid models (F. Caputo, private communication). 
Data listed in column 6 of Table \ref{distances} show that distances based on 
these theoretical NIR PL relations also agree within 1$\sigma$ with the BW estimates. 
Note that the distances and their uncertainties based on theoretical PL relations 
are larger than the empirical ones, since the former relations were derived 
including all the fundamental models with abundances ranging from $Z=0.001$ 
to $Z=0.04$ and different helium contents (Bono et al.\ 2010, in preparation).

\begin{table*}
\caption{Distances and radii according to the BW-analysis of the four selected 
Cepheids.}\label{distances} 
\begin{center}
\begin{tabular}{l c c r c c c c} \hline\hline
Name       &  $p$  &  $d$(pc)   &  $R/R_\odot$  &  $d$(pc)$_{PL}^{a}$  & $d$(pc)$_{PL}^{b}$& $(R/R_\odot)_{G07}^b$   & $(R/R_\odot)_{theo}^c$ \\ \hline
V340~Ara & 1.205 &3890$\pm$126$\pm$547& 100.2$\pm$3.2$\pm$17.3 &3154$\pm$686 & 3754$\pm$1156 &109.5$\pm$1.1 &115$\pm$1\\
UZ~Sct   & 1.217 &3176$\pm$116$\pm$152&  82.1$\pm$3.0$\pm$3.5 &3117$\pm$871 & 3540$\pm$1177 & 86.4$\pm$1.1 & 91$\pm$1\\
AV~Sgr   & 1.215 &2302$\pm$ 72$\pm$123&  93.4$\pm$2.9$\pm$4.2 &3036$\pm$792 & 3326$\pm$1063 & 89.2$\pm$1.1 & 94$\pm$1\\
VY~Sgr   & 1.219 &2546$\pm$ 53$\pm$134&  86.5$\pm$1.8$\pm$3.9 &2300$\pm$663 & 2533$\pm$ 849 & 81.1$\pm$1.1 & 86$\pm$1\\
\hline
\multicolumn{8}{l}{$^{\it{a}}$ Mean $JHK$-distances based on empirical NIR PL relations \citep{fo07}.} \\
\multicolumn{8}{l}{$^{\it{b}}$ Mean $JHK$-distances based on theoretical NIR PL relations
(Caputo, private communication).} \\
\multicolumn{8}{l}{$^{\it{c}}$ Radius estimate according to the empirical PR relation by G07.} \\
\multicolumn{8}{l}{$^{\it{d}}$ Radius estimate according to the theoretical PR relation by \cite{petr03}.} \\
\end{tabular}
\end{center}
\end{table*}

The mean Cepheid radii listed in column 4 of Table \ref{distances} agree quite well 
with mean radii based on empirical Period-Radius (PR) relation for Galactic Cepheids 
recently provided by G07 (see column 7 in Table \ref{distances}). Note that this PR 
relation is based on six Galactic Cepheids with known distances \citep{ben07} and with 
interferometrically measured angular diameter variations \citep{ker04b,mer05}. The same 
outcome applies in the comparison with radii based on the predicted PR relation for 
Galactic Cepheids provided by \cite{petr03}, assuming a solar chemical composition. 
Current radii agree on average within 10\% (two $\sigma$) with predicted and 
empirical PR relations. 

\section{Cepheid iron abundances}
To measure the iron abundances of the target Cepheids we followed the same 
approach suggested by \cite{bert07}. This approach relies on three different 
steps: 
\begin{itemize}
\item {\bf Line list}: The line list adopted by \cite{ro08}, includes 275 
\ion{Fe}{i} lines and 37 \ion{Fe}{ii} lines covering the FEROS spectral range. 
However, the target Cepheids are metal-rich and a good fraction of \ion{Fe}{ii} 
are saturated. In order to provide robust estimates of intrinsic parameters and 
in turn accurate measurements of iron abundances, we supplemented the line list 
by \cite{ro08} with 39 new weak \ion{Fe}{ii} lines (see Table \ref{lines}).
\item {\bf Equivalent width}: The measurement of the equivalent widths (EW) 
of the iron lines was performed using {\tt fitline}. This code uses a 
Gaussian fit, which is defined by four parameters: central wavelength, width, 
depth and continuum value of the individual lines. The initial value of the 
Gaussian parameters is fixed by randomly selecting the four parameters. Then, 
the ``genetic'' algorithm computes the $\chi^{2}$ between the observed line and the expected 
Gaussian profile and computes the new set of Gaussian parameters among the 20 
best fit solutions of the previous ``generation'' by applying random modifications 
of the values of the parameters (``mutation''). The algorithm after 100-200 
``generations'', gives the best fit Gaussian parameters (lowest $\chi^{2}$) 
for each observed line. For the measurement of the iron abundance, we have selected 
only lines with equivalent widths between 10 and 200 m$\AA$. The lower limit 
was chosen to be a safe compromise between the spectral characteristics and the 
need for weak lines for an optimal abundance determination. The upper limit was 
fixed to avoid the saturated portion of the curve of growth. 
\item {\bf Stellar Parameters}: The determination of an accurate effective 
temperature is a critical point in the abundance determination. This requirement 
becomes even more important if we are dealing with variable stars, since the 
temperature estimate has to refer to the pulsation phase at which the spectrum 
was collected. In this analysis, the effective temperature ($T_{\rm eff}$) was 
estimated spectroscopically using the line depth ratios (LDR) method described 
in \cite{kov00}. This technique has the advantage of being independent of 
interstellar reddening and minimally dependent on metallicity. These uncertainties 
plague other methods like the integrated flux method or the color-temperature 
relations \citep{gra94, kro98}. This approach becomes even more relevant for 
the target Cepheids, since they are located close to the edge of the inner 
Galactic disk, and therefore they are characterized by high reddening values 
(see column 5 in Table \ref{mag}). The estimated effective temperatures 
together with their errors are listed in columns 2 of Table \ref{iron}. 
The surface gravity ($\log~g$) and the microturbulent velocity ($v_{\rm t}$) 
were constrained by minimizing the log([Fe/H]) vs. EW slope (using the 
\ion{Fe}{i} abundance) and by imposing the ionization balance between 
\ion{Fe}{i} and \ion{Fe}{ii} (see column 3 and 4 in Table \ref{iron}). 
These two procedures are tightly connected and require an iterative process. 
The initial values  for the microturbulent velocity and the surface gravity, 
were fixed using typical Cepheid values ($v_{\rm t}$=$3~km~s^{-1}$, $\log~g$$=$2, 
\cite{and02a}). For the ionization balance, we assume that it was fulfilled 
when the difference between [\ion{Fe}{i}/H] and [\ion{Fe}{ii}/H] was smaller 
than the standard deviation on [\ion{Fe}{ii}/H] (typically, 
$\sigma_{[Fe/H]}\sim0.08-0.1$ dex). 
If this condition was satisfied by more than one value of $\log g$, we 
checked which value satisfies also the ionization balance within the 
standard deviation on [\ion{Fe}{i}/H] (typically, $\sigma_{[Fe/H]}\sim0.02$ dex). 
The effective temperatures listed in column 5 ($T_{\rm eff}$(Model)) are the 
final best fit values of the atmosphere models adopted in the iterative process.
To determine the errors on the microturbulent velocity and the surface gravity, we 
ran several iterations for each star, slightly modifying the values of 
these two intrinsic parameters that fulfill the requirements mentioned above. 
We have estimated that the intrinsic error on the microturbulent velocity is 
of the order of $0.1~km~s^{-1}$, while the intrinsic error on the surface 
gravity is of the order of $0.10$~dex. 
\end{itemize}

\begin{table*}
\caption{Intrinsic parameters and iron abundances for the target Cepheids.}\label{iron}
\begin{center}
\begin{tabular}{l l c c c c c c c} \hline\hline
Name&$T_{\rm eff}$(LDR)&$\log~g$&$v_{\rm t}$&$T_{\rm eff}$(Model)&[\ion{Fe}{i}/H]&[\ion{Fe}{ii}/H]&$<$[Fe/H]$>$&[Fe/H]$_{And}$\\\hline
V340~Ara &5425$\pm$146 &  0.2 & 3.5 &5475 &0.38$\pm$0.17 &0.42$\pm$0.26 & +0.40$\pm$0.21& +0.31$\pm$0.10 \\
UZ~Sct   &4790$\pm$95  &  0.6 & 3.7 &4850 &0.35$\pm$0.17 &0.35$\pm$0.27 & +0.35$\pm$0.21& +0.33$\pm$0.10 \\
AV~Sgr   &5407$\pm$53  &  0.9 & 4.2 &5450 &0.26$\pm$0.12 &0.28$\pm$0.12 & +0.27$\pm$0.12& +0.34$\pm$0.10 \\
VY~Sgr   &5268$\pm$146 &  1.0 & 3.2 &5400 &0.38$\pm$0.17 &0.36$\pm$0.26 & +0.35$\pm$0.21& +0.26$\pm$0.10 \\
\hline
\end{tabular}
\end{center}
\end{table*}

Our final \ion{Fe}{i} and  \ion{Fe}{ii} abundances, together with the adopted 
stellar parameters, are listed in Table \ref{iron}. Data listed in column 8 of 
this table indicate that the mean iron contents have weighted intrinsic 
uncertainties of the order of 0.20 dex, due to errors in the EW measurements, 
in the intrinsic parameters and in the number of unsaturated iron lines. 
This finding further supports the difficulty in measuring the iron 
abundance of metal-rich Cepheids. The intrinsic accuracy can be certainly 
improved using high S/N ratio, multi-epoch spectra of the same targets, since  
empirical evidence suggests that Cepheid elemental abundances minimally depend 
on the pulsation phase \citep{luck04, kov05}. 
Finally, it is worth mentioning that current iron abundances agree, within 
one $\sigma$, with the abundances provided by \cite{and02a} using a similar 
approach (see last column in Table \ref{iron}). 

\begin{figure}
\begin{center}
\includegraphics[width=7.0cm,angle=0]{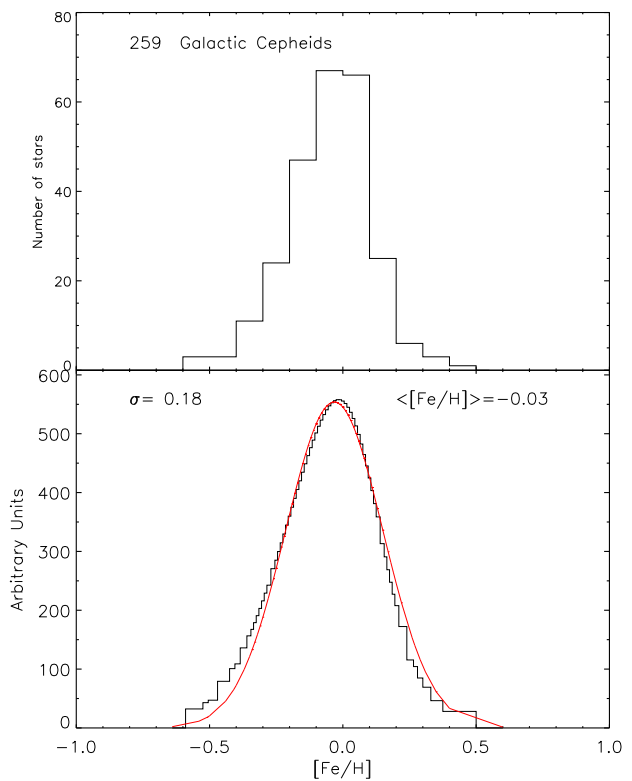}
\vspace{0.7cm}
\caption{Top -- Metallicity distribution of Galactic Cepheids. Bottom --
Same as the top, but the metallicity distribution was smoothed using a Gaussian 
kernel with standard deviation equal to the metallicity uncertainty of individual 
Cepheids. The red line shows the Gaussian fit of the metallicity distribution. 
The mean and the $\sigma$ are also labeled.}\label{isto}
   \end{center}
   \end{figure}

\section{Period-Amplitude diagram}

The metal-rich regime of classical Cepheids has been only marginally investigated, 
since till a few years ago iron measurements for these objects were not available. 
As we have already mentioned in \S 2, Walraven $VBLUW$ photometry is available 
for more than 160 Galactic Cepheids. The pulsation properties of these objects 
will be discussed in a forthcoming paper (Pedicelli et al.\ 2010). 
Interestingly, three out of the four target Cepheids belong to this 
sample, namely AV~Sgr, UZ~Sct and VY~Sgr. 
For these Cepheids were secured at least 30 phase points in 
five-$VBLUW$ bands that properly cover the entire pulsation cycle. 
The intrinsic accuracy of individual measurements is of the order of 
a few millimag. The uncertainty on the mean magnitudes, estimated 
using a fit with a cubic spline, is at most of the order of a few 
hundredths of a magnitude.
In order to investigate the possible correlation between luminosity amplitude 
and iron abundance, the  Walraven $B_W$,$V_W$-band photometric data were 
transformed into the Johnson $B_J$,$V_J$-band data using the relations:
 {\small
\begin{equation}
\label{vj}
V_J=6.886^{mag}-2.5 \; V_{W}-0.1916 \; (V-B)_{W}
\end{equation}
\begin{equation}
\label{bvj}
(B-V)_J=2.7947 \; (V-B)_{W}-1.2052  \; (V-B)_{W}^{2}+0.6422 \; (V-B)_{W}^{3}-0.0100
\end{equation}
} 

Accurate iron abundance measurements for Walraven Cepheids are available 
for 77 objects, while for 67 objects metallicity estimates are available 
based on the Walraven metallicity index \citep[][and references therein]{ped08}. 
This sample was supplemented with 115 Cepheids for which accurate iron abundances 
exist in the literature, based on high resolution spectra 
\citep{and02b, and02a, and02c, and04, bert07, szi07, bert08, ro08}.
We ended up with a sample of 259 Galactic Cepheids and the top panel of Fig. 
\ref{isto} shows the metallicity distribution. In order to provide accurate 
estimates of the mean metallicity and of the spread in metallicity of the 
Galactic disk, we ran a Gaussian kernel with a $\sigma$ equal to the 
metallicity uncertainty of individual Cepheids. The metallicity distribution 
we obtained is plotted in the bottom panel of Fig. \ref{isto} and the red line 
shows the Gaussian fit. We found a solar mean metallicity ($<[Fe/H]>\sim$ -0.03) 
and a sigma of 0.18 dex. These values agree quite well with similar metallicity 
distributions available in the literature \citep{chiap01, ces07}. 
Our metallicity distribution is also asymmetric, and 
indeed the metal-poor tail is shallower than the metal-rich one. 
This finding agrees with predictions based on Galactic chemical evolution 
models \citep{holm07, yin09}.

\begin{figure}
\begin{center}
\includegraphics[width=7.0cm,angle=0]{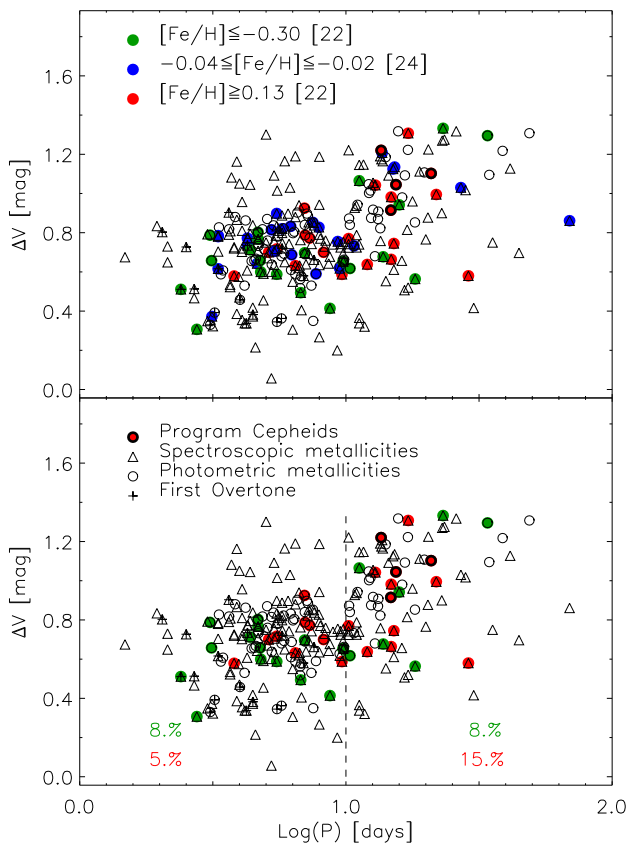}
\vspace{0.7cm}
\caption{Top -- $V$-band amplitude versus period for a sample of 259 Galactic 
Cepheids. Triangles and circles display spectroscopic and photometric metallicities, 
respectively. Red and green dots mark metal-rich ([Fe/H]$\geq 0.13$) and 
metal-poor ([Fe/H]$\leq$ -0.30) Cepheids, while the blue ones show Cepheids 
with iron abundances ranging from -0.04 to -0.02 dex. The selected metal-rich 
Cepheids are marked with a cross, while the pluses mark first overtone pulsators. 
Bottom -- same as the top, but the vertical dashed line splits short-  ($\log(P)\le$ 1.0) 
and long-period ($\log(P)>$ 1.0) Cepheids. The fractions of metal-poor and 
metal-rich Cepheids are also labeled.}\label{PAV}
\end{center}
\end{figure}

\begin{figure}
\begin{center}
\includegraphics[width=7.cm,angle=0]{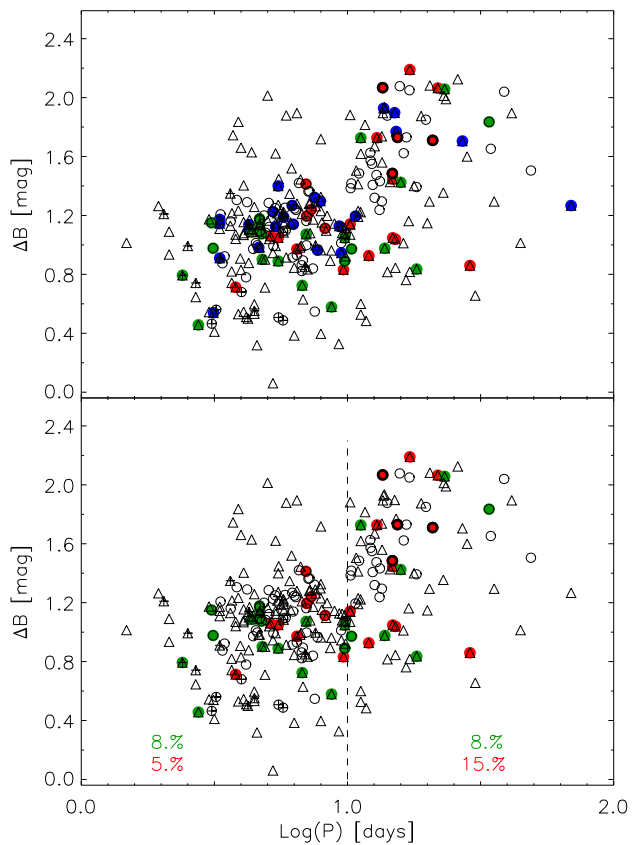}
\vspace{0.7cm}
\caption{Same as Figure \ref{PAV}, but for the $B$-band amplitudes.}\label{PAB}
   \end{center}
   \end{figure}

To constrain the dependence of the pulsation properties on the metal content, 
we adopted the Bailey diagram, i.e. luminosity amplitude vs.  pulsation 
period. Data plotted in Figures \ref{PAV} and \ref{PAB}  show the V, B 
luminosity amplitudes of the Cepheid sample adopted by 
Pedicelli et al. (2009). The amplitudes are based
either on Walraven photometry --transformed into the Johnson system 
(see \S2.1)--  or on data in the Galactic Cepheid catalog provided 
by \cite{fer95}.  To constrain the possible dependence of the pulsation 
amplitude on the metal content we selected three different sub-samples 
representative of the metal-poor ([Fe/H]$\le$-0.30 dex) tail, of the 
metal-rich ([Fe/H]$\ge$0.13 dex) tail and of the peak metallicity 
(-0.04 $\le$[Fe/H]$\le$-0.02 dex).   
The three samples roughly include two dozen of Cepheids. Data plotted in the top 
panels of Figures \ref{PAV} and \ref{PAB} display that both $V$ and $B$-band 
amplitudes are not correlated with the metal content. This feature does not 
agree with a well established theoretical and 
empirical evidence of RR Lyrae stars \citep{bo07}, i.e. the prototype 
of low-mass, helium burning radial variables located inside the so-called 
Cepheid instability strip. Such a difference can be 
partially explained with the empirical circumstance that Galactic RR Lyrae 
variables cover almost three dex in metal content, when moving from the halo to 
the bulge, while Galactic Cepheids in the disk only cover one dex. 
Moreover, the Hertzsprung progression causes a systematic decrease in the 
pulsation amplitudes for periods across ten days \citep{payne51, payne54}. 
Current empirical \citep{andrea87, welch97, beau98, mos00} and 
theoretical \citep{bo00b}  
evidence indicates that a decrease in metal content causes a systematic drift 
in the center of the Hertzsprung progression toward shorter periods. This means 
a reshuffle in the pattern of the luminosity amplitudes as a function of the 
pulsation period when dealing with Cepheids with different metal abundances.   
However, the spread in the amplitudes is also present at periods shorter 
and longer than the Hertzsprung progression. This indicates that the lack of 
a well defined correlation with metallicity might also be due to 
the dependence of the topology of the instability strip on the chemical 
composition \citep{pellub78, bo99, bo00a} to evolutionary effects and to binarity 
(see Pedicelli et al.\ 2010, in preparation).    
    
As a final test concerning the metallicity distribution of Galactic Cepheids we 
split the sample into short ($\log P \le 1.0$) and long ($\log P > 1.0$) period 
objects. Data plotted in the bottom panels of Figures \ref{PAV} and \ref{PAB} 
show that metal-rich ([Fe/H]$\ge$-0.13 dex) Cepheids are more frequent  
among long- than among short-period Cepheids (15\% vs 5\%) Cepheids. 
On the other hand, the metal-poor  ([Fe/H]$\le$-0.30 dex) Cepheids are equally 
distributed between the two groups. The evidence that the period distribution of 
classical Cepheids depends on the metal-content dates back to 
\citet[][and references therein]{gas74}, 
who realized that the peak shifts toward shorter periods when moving from 
Galactic to Small Magellanic Cloud Cepheids. 
Subsequent evolutionary tracks for intermediate-mass stars showed that more metal-poor 
structures are characterized, at fixed mass, by blue loops that cover a larger 
temperature range. This means that the minimum Cepheid mass crossing the instability 
strip is smaller in metal-poor that in metal-rich systems. As a consequence a 
systematic drift in the period distribution of metal-rich Cepheids is expected. 
However, the period cut we adopted is well beyond the short period cutoff of 
Galactic Cepheids, thus suggesting that the paucity of metal-rich Cepheids 
among short-period Cepheids is probably due to an observational bias.       

\section{Summary and conclusions}

We provided accurate BW distances and radii for four metal-rich Cepheids, namely 
V340~Ara, UZ~Sct, AV~Sgr and VY~Sgr. Current distance estimates, taken at face value, 
agree quite well with similar estimates based either on empirical \citep{fo07} or on 
theoretical NIR PL relations. However, the uncertainties affecting the BW distances, 
by summing in quadrature the errors in the fit and the errors estimated with the Monte 
Carlo simulations, are on average a factor of 3-4 smaller than for distances based on 
predicted and empirical NIR PL relation. The same outcome applies to the mean 
Cepheid radii, but the uncertainties on the BW radii, by summing in quadrature errors 
on the fit and errors from Monte Carlo simulations, are on average a factor 
of two larger than for radii based either on the empirical or on the theoretical 
PR relation provided by G07 and by \cite{petr03}, respectively.  

We also collected high-resolution, high signal-to-noise ratio spectra to measure 
the iron abundances of the target Cepheids. Special attention was paid to 
provide accurate estimates of intrinsic parameters (effective temperature, 
surface gravity, microturbulent velocity) directly from observed spectra. 
We performed detailed measurements of iron abundances using large samples 
of \ion{Fe}{i} and \ion{Fe}{ii} lines. Current abundances indicate that selected 
Cepheids are super metal-rich and agree, within 1$\sigma$, with iron abundances 
provided by \cite{and02a} using a similar approach. 

We adopted a sample of 259 Galactic Cepheids for which are available either 
spectroscopic iron measurements or metallicity estimates based on the 
Walraven metallicity index \citep{ped08}. We found that classical 
Cepheids do not seem to show in the Bailey diagram (luminosity amplitude 
vs pulsation period), in contrast with low-mass helium burning RR Lyrae stars, 
a clear correlation between luminosity amplitude and metallicity. The lack of 
such a correlation might be the consequence of the Hertzsprung progression.  
We also found that a good fraction of metal-rich ([Fe/H]$\ge$0.13 dex) Cepheids 
are located among long-period ($\log P \ge$ 1.0) variables. However, for the 
moment this can only be considered as circumstantial evidence, since the current 
sample is probably affected by selection bias. 
Metal-rich Cepheids are located in the inner disk and are typically affected 
by large extinctions. The selected Cepheids have Galactocentric distances 
smaller than 6.5 kpc and their reddening ranges from 0.6 to 1.3 mag. Detailed analysis 
concerning the pulsation properties of metal-rich Cepheids best awaits more complete 
samples.  However, the game is worth the candle, since classical Cepheids are excellent 
tracers of young stellar populations. Their pulsation properties and their radial distribution 
across the inner Galactic disk and the bar can provide robust constraints on the bar-driven formation 
scenario \citep{vanloo03, deba04, zoc06} on short (10-100 Myr) timescales.

\begin{acknowledgements}
We acknowledge an anonymous referee for his/her positive opinion
concerning the content of this investigation. Two of us (SP, GB) 
acknowledge ESO support (DGDF funds) for a stay in Garching, 
during which a good part of this paper was written. It is a pleasure 
to thank M. Zoccali 
for interesting discussions concerning the interaction between the Galactic disk and the 
bulge. We also thank G. Iannicola and I. Ferraro for their suggestions concerning 
Gaussian smoothing.   
\end{acknowledgements}

\bibliographystyle{aa}
\bibliography{mybib}

\end{document}